\documentclass[conference]{IEEEtran}
\IEEEoverridecommandlockouts
\usepackage{cite}
\usepackage{amsmath,amssymb,amsfonts}
\usepackage{algorithm}
\usepackage{algorithmic}
\usepackage{graphicx}
\usepackage{textcomp}
\usepackage{xcolor}
\def\BibTeX{{\rm B\kern-.05em{\sc i\kern-.025em b}\kern-.08em
    T\kern-.1667em\lower.7ex\hbox{E}\kern-.125emX}}
\begin{document}

\title{A Distributed Multi-GPU System for Large-Scale Node Embedding at Tencent
}

\author{\IEEEauthorblockN{1\textsuperscript{st} Wanjing Wei}
\IEEEauthorblockA{\textit{WeChat} \\
\textit{Tencent Inc.}\\
Guangzhou, China \\
aaronwei@tencent.com}
\and
\IEEEauthorblockN{2\textsuperscript{nd} Yangzihao Wang}
\IEEEauthorblockA{\textit{WeChat} \\
\textit{Tencent Inc.}\\
Beijing, China \\
slashwang@tencent.com}
\and
\IEEEauthorblockN{3\textsuperscript{rd} Pin Gao}
\IEEEauthorblockA{\textit{WeChat} \\
\textit{Tencent Inc.}\\
Shenzhen, China \\
powergao@tencent.com}
\and
\IEEEauthorblockN{4\textsuperscript{th} Shijie Sun}
\IEEEauthorblockA{\textit{WeChat} \\
\textit{Tencent Inc.}\\
Shenzhen, China \\
cedricsun@tencent.com}
\and
\IEEEauthorblockN{5\textsuperscript{th} Donghai Yu}
\IEEEauthorblockA{\textit{WeChat} \\
\textit{Tencent Inc.}\\
Shenzhen, China \\
hunteryu@tencent.com}
}

\maketitle

\begin{abstract}
  Real-world node embedding applications often contain hundreds of billions of edges with high-dimension node features. Scaling node embedding systems to efficiently support these applications remains a challenging problem. In this paper we present a high-performance multi-GPU node embedding system. It uses model parallelism to split node embeddings onto each GPU's local parameter server, and data parallelism to train these embeddings on different edge samples in parallel. We propose a hierarchical data partitioning strategy and an embedding training pipeline to optimize both communication and memory usage on a GPU cluster. With the decoupled design of CPU tasks (random walk) and GPU tasks (embedding training), our system is highly flexible and can fully utilize all computing resources on a GPU cluster. Comparing with the current state-of-the-art multi-GPU single-node embedding system, our system achieves 5.9x---14.4x speedup on average with competitive or better accuracy on open datasets. Using 40 NVIDIA V100 GPUs on a network with almost three hundred billion edges and more than one billion nodes, our implementation requires only 3 minutes to finish one training epoch.
\end{abstract}

\section{Introduction}
\label{sec:intro}
%Motivations
Network data structure has been widely used to present complex relationships between entities in various fields such as social networks, finance, e-commerce, and molecular biology. To use network data structure in machine learning, a common and important method is node embedding, which maps each node into a lower dimension vector space where its embedding can be learned and served as features for downstream machine learning tasks. Nowadays, with both network data and embedding features getting larger and larger, there is a rising demand for the efficient embedding training of large-scale networks such as online social networks~\cite{Lerer:2019:PBG} or e-commerce recommender systems~\cite{Wang:2018:BSC}.

%Challenges
Heterogeneous node embedding systems with both CPUs and GPUs~\cite{Zhu:2019:GVA} improve the performance of distributed CPU node embedding systems~\cite{Lerer:2019:PBG,Tang:2015:LINE}, but can only provide solutions for networks at moderate scale. Distributed CPU node embedding systems, on the other hand, are able to handle large-scale networks, but fail to deliver high performance with GPU computing. Supercomputers with GPUs provide high training performance mainly due to specialized communication channel and topology design that cannot apply to most GPU clusters with uneven speed of communication at different hierarchical level and restrained on device memory~\cite{Yang:2021:EAE}. How to build a distributed node embedding system on a commodity GPU cluster that can both scale to billion-level node trillion-level edge networks and deliver high performance is still a challenge, due to the data and model scale.

%More specifically, there are several research questions:
%\begin{enumerate}
%    \item For tasks with both large-scale data and model, such as node embedding, neither pure data parallelism nor pure model parallelism training could scale well to distributed clusters, how can one design a parallel training method to enable scalable training on distributed clusters?
%    \item With number of edges at hundred-billion scale, how can one optimize memory-bound operations with unique irregular memory access patterns to gain high performance?
%    \item With both data storage and computation in a distributed fashion, inter-GPU, CPU-GPU, and inter-node communication are all in large volumes. How can one design communication strategy to reduce communication cost?
%    \item Node embedding task combines network-oriented primitives and machine learning-oriented primitives. How to make them collaborate efficiently?
%\end{enumerate}

%Our Contributions
In this paper we propose a high-performance node embedding system that runs on a GPU cluster. Our contributions are as follows:
\begin{itemize}
    \item We design and implement a high-performance large-scale node embedding system that uses hybrid model data parallel training and can run on any commodity GPU clusters;
    \item We propose a hierarchical data partitioning strategy as well as a group of optimizations to enable high-performance training of the embeddings;
    \item We propose a pipelined embedding training strategy that enables our system to efficiently overlap computation and communication to achieve both high throughput and high communication bandwidth;
    \item We design our node embedding system to decouple the random-walk-based network augmentation stage with the embedding training stage so that different optimization strategies could be applied for each component to make the full task run efficiently.
\end{itemize}
On several models with real-world networks, our system shows excellent scalability and achieves an order of magnitude speedup on average over the current state-of-the-art GPU network embedding framework without losing accuracy.

The rest of the paper goes as follows: Section~\ref{sec:pre} provides preliminary knowledge that includes node embedding and network partitioning. Section~\ref{sec:overview} gives an overview of our system as well as several key strategies we propose: hybrid data model parallel training, hierarchical data partitioning, and pipelined embedding training. Section~\ref{sec:impl} provides implementation details of the system. Section~\ref{sec:exp} provides extensive performance analysis based on our experiment results. Section~\ref{sec:related} discusses the related works and Section~\ref{sec:conc} concludes with future work.

\section{Preliminaries}
\label{sec:pre}
\subsection{Node Embedding Methods}
Given a network $G=(V, E)$, where $V$ is the set of vertices and $E$ is the set of edges. Node embedding refers to the approach of learning latent low-dimensional feature representations for the nodes in the network, with the goal of capturing the structure of the original networks without relying on the inflexible and expensive hand-engineered features.

\textbf{Training Data:} Due to the sparse nature of many real-world networks, using only real edges as positive samples usually cannot achieve acceptable training results. Hence most existing embedding methods use various \textit{network augmentation} methods to generate additional positive samples. The most popular ones are random-walk-based methods: Each node in the original network generates random walk paths and uses nodes on the paths to form additional positive samples. To increase the model's discriminative power, most existing embedding methods also use negative sampling to provide negative samples. The training objective is to distinguish the positive samples after network augmentation from negative samples.

\textbf{Embedding Representation:} Embeddings are usually encoded in two sets: \textbf{vertex} embedding matrix and \textbf{context} embedding matrix since one node can be treated as either the source node or the destination node when shared among more than one edge. 

\textbf{Training Process:} embedding training on the augmented network involves picking a batch of samples with both positive samples and negative samples, computing the similarity score of vertex[u] and context[v] for an edge sample (u, v), and optimizing (e.g. via standard SGD) to encourage neighbor nodes to have close embeddings, whereas distant nodes to have very different embeddings. Algorithm~\ref{algo:emb} shows the pseudo code. The training process takes several epochs, where one epoch goes over all the sample edges in the augmented network, and can be broken up into several episodes where each episode train embeddings on one part of sample edges.

\begin{algorithm}
 \caption{Random walk-based node embedding}
 \begin{algorithmic}[1]
 \label{algo:emb}
 \renewcommand{\algorithmicrequire}{\textbf{Input:}}
 \renewcommand{\algorithmicensure}{\textbf{Output:}}
 \REQUIRE $G=(V, E)$, embedding dimension $d$, walk distance $k$, walk context length $l$
 \ENSURE  vertex embeddings $Emb_{vertex} \in R^{|V| \times d}$
 \\ \textit{network augmentation} :
  \STATE $E_{aug} := E$
 \\ \textit{parallel random walk}
  \FOR {$v \in V$}
    \FOR {$u \in walk(v, k, l)$}
      \STATE $E_{aug} := E_{aug} \cup (v, u)$ 
  \ENDFOR
  \ENDFOR
\\ \textit{embedding training} :
\FOR {each iteration}
  \STATE $v,u := EdgeSample(E_{aug})$
  \STATE $Train(Emb_{vertex}(v), Emb_{context}(u), label=1)$
  \FOR {$u' \in NegativeSample(E_{aug})$}
    \STATE $Train(Emb_{vertex}(v), Emb_{context}(u'), label=0)$
  \ENDFOR
\ENDFOR
 \RETURN $Emb_{vertex}$
 \end{algorithmic}
 \end{algorithm}

\subsection{Network Partitioning}
To handle networks with billions of vertices and hundred billion of edges on a GPU cluster, we have to partition it across different nodes. Given $k$ nodes and a network $G=(V,E)$, network partitioning tries to divide $G$ into $k$ subgraphs ${G_i = (V_i, E_i) | 1 \le i \le k}$, such that $V_1 \cup V_2 \cup ... \cup V_k = V$ and $E_1 \cup E_2 \cup ... \cup E_k = E$. One can use either 1D partitioning or 2D partitioning for different types of applications. 

%1D partitioning method is vertex-centric. It contains two common methods: \textbf{Edge-Cut} method first divides the vertex set $V$ into $k$ subset, then for edges that connect vertices from different subset, it creates local mirror vertices and add edges to connect these mirror vertices to the original vertices. \textbf{Vertex-Cut} method also divides the vertex set, then for edges share one same vertex but have the other vertex belong to different subsets, it adds duplicated vertices.

%\begin{figure}[htbp]
%  \centerline{\includegraphics[width=\linewidth]{figures/partition.pdf}}
%  \caption{For a cluster with $k$ processing units, 2D partitioning splits the edge set into $k^2$ subsets, where subset $E_{i,j}$ contains edges that have source node from $V_i$ and destination node from $V_j$. In this way, our context and vertex embeddings are partitioned across processing units (in our system they are GPUs), and edge samples are partitioned by both source and destination nodes. By properly designing which blocks to train in parallel, 2D partition guarantees that edge samples that belong to different partitions have orthogonal vertex usage.}
%  \label{fig:partition}
%\end{figure}

Since both random walk-based network augmentation and embedding training are edge-centric, in this work we use 2D partitioning method. To train a batch of edge samples, one GPU needs data from three sources: \textbf{vertex embeddings} for source nodes, \textbf{context embeddings} for destination nodes, and \textbf{sample pool} that contains edges in the form of source node ID and destination node ID pairs. As shown in Figure~\ref{fig:hybrid}, in one round of training using a GPU cluster with two 8-GPU computing nodes, edge samples with different source node ID range and destination node ID range are loaded from sample pool. Vertex embeddings and context embeddings are transferred among GPUs and nodes, so that one part of embeddings get trained on all samples that share source node IDs with it.

%\subsection{Performance Modeling}
%Table~\ref{tab:mem} shows the memory cost for network topology data and embedding data for our system. To model the performance of node embedding task, without losing generality, we limit the training process to be within one epoch. For a network with $n$ samples and nodes with $d$ dimensions, the memory complexity is $O(nd)$ which includes loading both vertex and context embeddings for each edge sample and updating the embeddings after the back propagation. The computation complexity is also $O(nd)$ which includes $d$ fused multiply-add operations for each edge pair and roughly two times of the workload for back propagation. While models for computer vision (e.g. ResNet-50~\cite{He:2015:DRL}) or natural language processing (e.g. BERT~\cite{Devlin:2019:BERT}) that involve more complex computations per sample usually have hundreds or thousands times of arithmetic intensity, our analysis shows that node embedding training has an $O(1)$ arithmetic intensity, which makes it a memory bound problem. Due to the heavy memory cost and relatively low arithmetic intensity, we focus on optimizing memory and communication, which serve as two critical factors for the high performance of our node embedding system.

%system overview illustration.
\begin{figure*}[htbp!]
  \centerline{\includegraphics[width=0.7\linewidth]{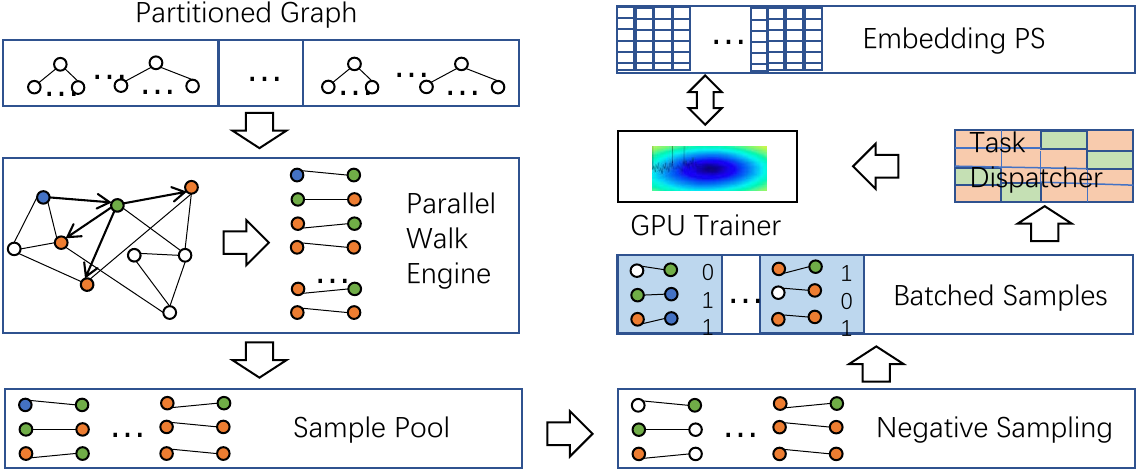}}
  \caption{Illustration of our node embedding system.}
  \label{fig:sys}
\end{figure*}

\section{Scaling Node Embedding on A GPU Cluster}
\label{sec:overview}
Our node embedding system consists of two major components (Figure~\ref{fig:sys}), namely parallel walk engine and embedding training engine. Parallel walk engine uses partitioned graph data to generate samples to sample pool (which can be either on disk or in memory). Batched with samples generated by negative sampling, these training samples are sent to a task dispatcher, which loads orthogonal samples to GPU trainers for training the embeddings. The system also contains an on-device embedding parameter server (PS) that manages node embeddings on each GPU. It is built specifically for large-scale random-walk based node embedding algorithms, which are the most common choices for existing node embedding systems. To handle the challenges of both huge model/data usage and heavy inter-node communication, we design a hybrid model data parallel training method. It uses a hierarchical data partitioning strategy and a pipeline design to optimize both memory and communication. It also decouples the random-walk/sampling phase with embedding training phase so that the system is scalable to huge networks with both the random walk part and the embedding training part running efficiently.

\begin{table}[htbp]
\caption{Memory cost of our node embedding system for a typical large-scale social network.}
\begin{center}
\begin{tabular}{|c|c|c|c|}
\hline
\textbf{Data}&\multicolumn{3}{|c|}{\textbf{Metrics}} \\
\cline{2-4} 
\textbf{Type} & \textbf{\textit{size}}& \textbf{\textit{example}}& \textbf{\textit{storage}} \\
\hline
nodes  & $|V|$ & 1.05 Billion & 3.91GB \\
edges  & $|E|$ & 300 Billion & 2.24TB \\
augmented edges & $|E'|$ & 3 Trillion & 22.4TB \\
vertex embeddings & $|V| \times d$ & 1.05*$10^{9}$*128 & 500.7GB \\
context embeddings & $|V| \times d$ & 1.05*$10^{9}$*128 & 500.7GB \\
\hline

\end{tabular}
\label{tab:mem}
\end{center}
\end{table}

For random-walk based node embedding systems, running random walk on a hundred billion edge level network needs several terabytes of memory and thousands of CPU cores; meanwhile, training the embeddings on GPUs requires the amount of GPU device memory that only a multi-node multi-GPU cluster can provide (Table~\ref{tab:mem}). With a typical heterogeneous setting that has eight GPUs on one CPU node, it is impossible to meet the requirements of both random walk and embedding training by directly migrating any system designed for single-node to our scenario. Thus, we decouple the walk engine and the embedding training engine. In our system, random walk and embedding training are two stages that run asynchronously. Random walk engine writes walks to sample pool, embedding training engine sends samples and embeddings to GPU and performs training. Random walk takes place on CPU cores, while embedding training happens on GPUs. This way we can adjust the paces of random walk producing and consuming by tuning both engines to maximize the performance. With such a design as our guideline, we have proposed a set of strategies to make our system run efficiently.

\subsection{Hybrid Data Model Parallel Training}

Data parallelism has been widely used as a common training method for distributed training of large-scale machine learning models. However, one limitation of data parallelism is the requirement of having a model size that can fit into the memory of the training unit, if the dataset size and the model size both exceed the memory limitation for one node, it needs to be combined with model parallelism. On a cluster with $N$ nodes and $M$ GPUs on each node, we first use 2-D partition strategy to partition edge samples (augmented and with negative sampling) into $(N*M)^2$ sample blocks. We use data parallelism to send batches of edge samples to all $N*M$ GPUs on the cluster from sample pool, and we use model parallelism to partition the node embeddings and send each partition to the embedding PS which is distributed on GPUs. During the training of one epoch, we locally train model with different embeddings in parallel for $N*M$ episodes, where after each episode we transfer each partition of the embeddings to a new GPU with a different edge sample block using multi-level ring-based communication. In this way, a single embedding will eventually see all edge samples that include it. Our hierarchical data partitioning strategy and embedding training pipeline design can hide most synchronization costs.

%hybrid training illustration.
\begin{figure*}[htbp!]
  \centerline{\includegraphics[width=\linewidth]{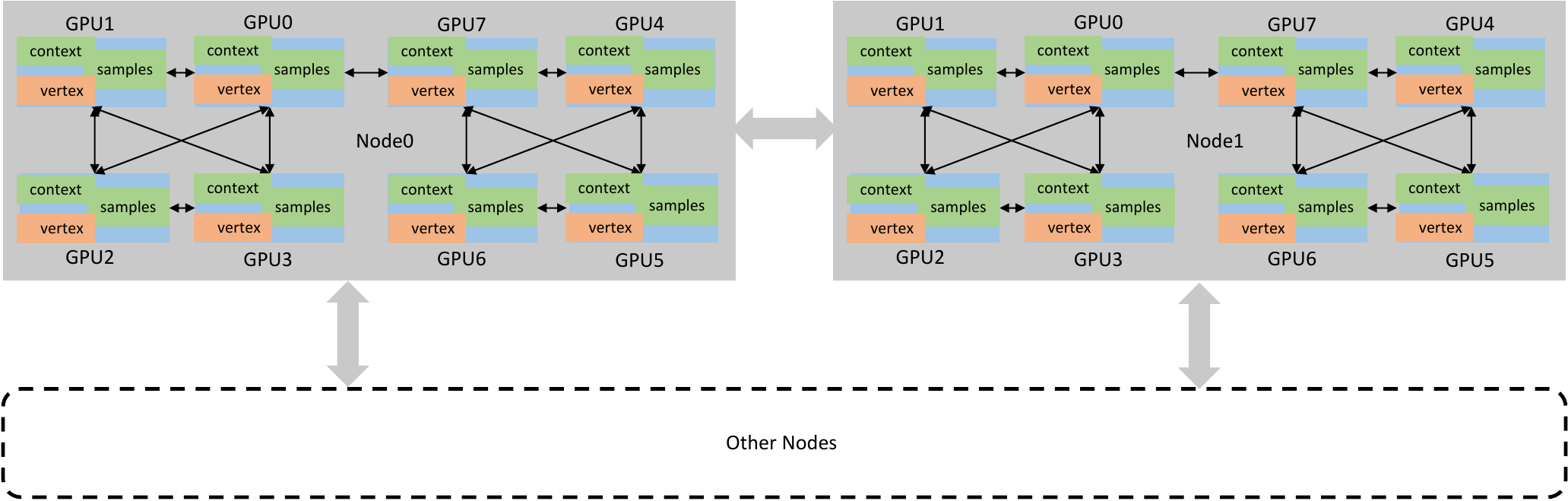}}
  \caption{Our hybrid model data parallel training uses data parallelism (light green blocks) to train all edge samples in multiple episodes by splitting samples for each episode onto local GPUs. It uses model parallelism (light orange blocks) to split vertex embeddings onto local GPUs and does two-level (inter-node and inter-GPU intra-node) ring-based communication during the training of each episode.}
  \label{fig:hybrid}
\end{figure*}

Splitting edge samples for data parallel training is easy as it can be done as a pre-training step with simple partitioning strategy. While partitioning the model is difficult as each part of the model needs to be trained with all edge samples, which are distributed over all GPUs in the cluster. Thus, we need to design a method that enables efficient data parallel training as well as efficient intra-node and inter-node communication.

\subsection{Hierarchical Data Partitioning}

GraphVite \cite{Zhu:2019:GVA} uses a 2D-partitioning strategy to divide samples into $n \times n$ blocks where $n$ is the number of GPUs on one node. All $n$ GPUs then take orthogonal blocks from samples and train local vertex embeddings before they load a new set of orthogonal blocks and transfer vertex embeddings using ring-based communication. 

\begin{figure}[htbp]
  \centerline{\includegraphics[width=0.7\linewidth]{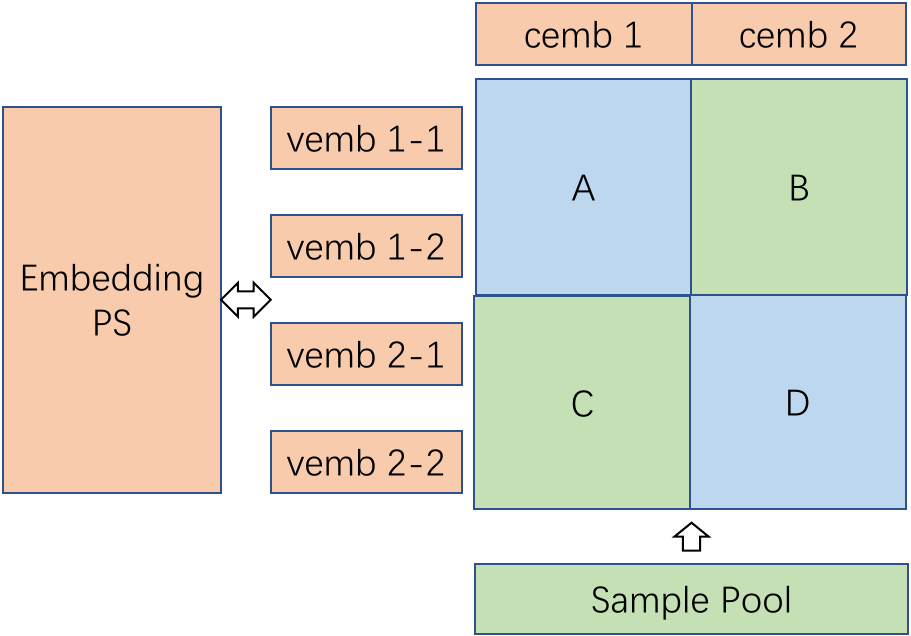}}
  \caption{Our intra-node embedding data partitioning strategy. vemb i-j holds the jth subpart of vertex embedding on the ith GPU. A,B,C,D are subparts of edge samples, where A and D locate on GPU1, B and C locate on GPU2.}
  \label{fig:sub_partition}
\end{figure}

However, it is impossible to directly apply GraphVite's strategy to a multi-node multi-GPU cluster where inter-node communication behaves very different from intra-node inter-GPU communication. Nowadays, this gap between inter-node and intra-node inter-GPU communication are getting even larger due to the ever-increasing inter-GPU communication speed enabled by NVLink. A multi-node multi-GPU system must consider this difference. We developed the hierarchical data partitioning strategy specifically for multi-node multi-GPU scenario. To handle two parts of the model, namely vertex embeddings and context embeddings, we can choose to let our on-device embedding PS store both embeddings, and each round transfer partitions of both context embeddings and vertex embeddings. 
Since this will make our partitioning strategy and pipeline design more complicated, we instead choose to fix the context embeddings for each GPU so that it could be loaded once and stay on device, which reduces half of our communication cost on embeddings. For vertex embeddings that need to be trained with different edge sample blocks on different GPUs, we hierarchically partition the full vertex embedding matrix: first inter-node, then intra-node and inter-GPU. Figure~\ref{fig:hybrid} illustrates our strategy on a cluster with two nodes where each node has eight GPUs: We first partition all the samples of this episode into sixteen blocks, eight on Node0 and eight on Node1. At the inter-node level, all GPUs from Node0 will first train on half of the vertex embeddings, then we use inter-node communication to swap the vertex embeddings for each node before we train the second half of the vertex embeddings. At intra-node inter-GPU level, we use an improved orthogonal block training strategy to guarantee that all available GPUs in the cluster train in parallel in fully synchronous style, which usually leads to better and faster convergence~\cite{dutta:2018:STALEGRAD}.

To fully utilize both communication bandwidth and computing resources, we further split vertex embeddings on one GPU into $k$ sub-parts. We tradeoff between memory usage and computation speed by using properly designed ping-pong buffers on GPUs. Figure~\ref{fig:sub_partition} shows an example inside one node with two GPUs where $k$ is set to 2. In our design, while one sub-part of the vertex embeddings are getting trained on each GPU, the other(s) are either pushing embeddings back to CPU memory (Device to Host communication) or pulling from CPU memory (Host to Device communication), which utilize both copying streams on the GPU~\@. By breaking vertex embeddings and edge samples into fine-grained sub-parts, we have more flexibility by tuning the size and the number of sub-parts on one GPU to get better performance. The ping-pong buffer design also allows us to pipeline our computation with communication to maximize the end-to-end training performance. Here the only stall stage is inter-GPU peer-to-peer communication stage, but because we only need to send one sub-part of our local vertex embeddings, the communication cost for one inter-GPU transfer is cut to $1/k$ when comparing with the naive data partitioning strategy. The choice of $k$ is critical to the performance. Setting the sub-part number too large will reduce the communication cost, but also reduce the computation workload for each sub-part, setting the number too small will increase the communication cost and make it difficult to be hidden by computation. In our implementation we have carefully tuned the number of $k$ to be equal to four since we found that it works the best on all our tasks and GPU clusters with various hardware settings.

Our design could be extended and scale to networks with various sizes. It adapts to both flexible number of GPUs on one node and flexible number of nodes on one cluster.

\begin{figure*}[htbp]
  \centerline{\includegraphics[width=0.8\linewidth]{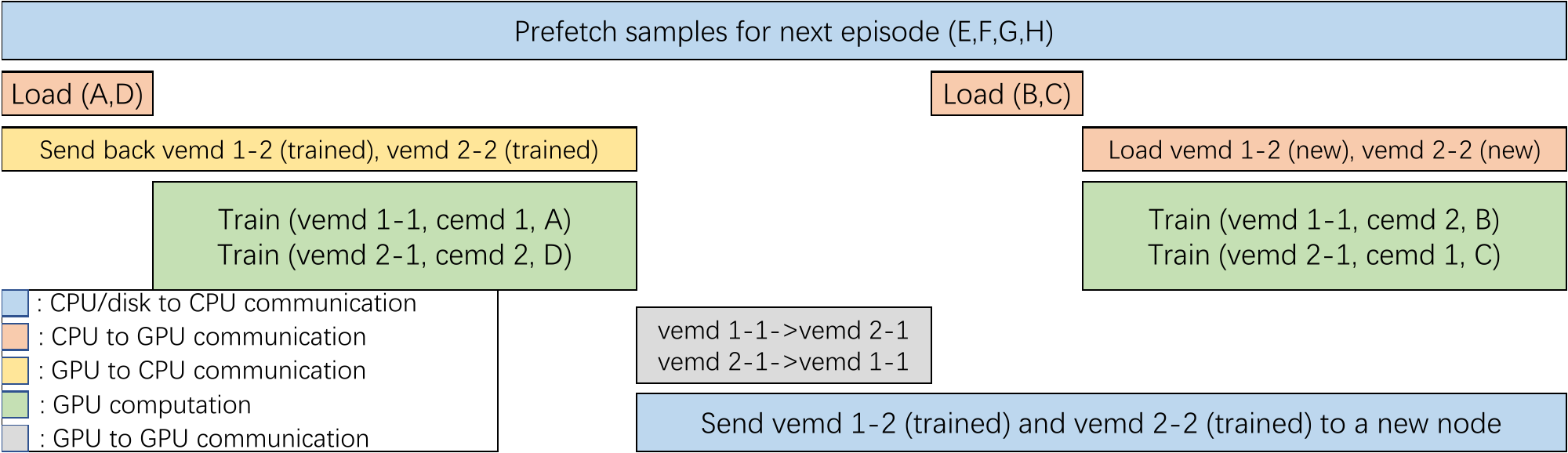}}
  \caption{The Pipeline (in the form of a timeline from left to right) of one episode of embedding training in our system using Figure~\ref{fig:sub_partition}'s settings as an example. Blocks that are vertically arranged in parallel run in parallel. Note that this is only a figurative illustration that shows components in the embedding training pipeline, the length of each block and its run time are not in proportion.}
  \label{fig:pipeline}
\end{figure*}

\subsection{Embedding Training Pipeline}

As we have analyzed in the previous section, a na\"ive implementation of data partitioning on multi-node multi-GPU system divides vertex embeddings into $N$ parts, where $N$ is the number of GPUs in the system. Each node uses orthogonal block training method and ring-based communication to train all local vertex embeddings on all local GPUs. Then inter-node ring-based communication happens to load a new set of vertex embeddings onto each node for training. Without a pipeline and ping-pong buffer design, this method will not be able to properly overlap communication with computation at both intra-node and inter-node levels.

Using our hierarchical data partitioning strategy, one epoch of embedding training involves various types of communication:
\begin{itemize}
  \item{Disk to CPU Memory:} Load random walks from file, generate and send edge samples to all node's memory;
  \item{Device to Host:} Transfer vertex embeddings that have finished training on all GPUs on the current node from GPU memory to CPU;
  \item{Host to Device:} Load vertex embeddings just arrived from other node from CPU to GPU memory, load a batch of edge samples from CPU to GPU memory;
  \item{Device to Device:} Send and receive vertex embeddings to train from other GPU;
  \item{Inter-Node:} Transfer trained vertex embeddings to other nodes.
\end{itemize}

To maximize the overlap between computation and communication, we propose an embedding training pipeline. As shown in Figure~\ref{fig:pipeline} (for the convenience of illustration, here we assume our cluster has two GPUs on one node), a typical pipeline of embedding training contains the following steps:
\begin{itemize}
    \item [-] While sample block A,B,C,D are currently being used, a thread starts pre-fetching sample blocks E,F,G,H which will be used for next episode;
    \item [-] Training phase starts with loading sample blocks to the GPU. Here we load A and D first; Then use local context embedding and these two sample blocks to train local vertex embedding for the current round in the episode;
    \item [-] While training, each GPU will send back one sub-part of local vertex embeddings which have finished training on all GPUs back to CPU memory. Note that after each round in the episode, there will be one more sub-part of vertex embeddings on each GPU that finishes training and is ready to be sent to other node;
    \item [-] When one sub-part finishes its training on the GPU, we use peer-to-peer communication between GPUs to send this sub-part of vertex embeddings to a new GPU with a different sample block;
    \item [-] While a new round of training is going on, the embedding PS on each GPU will load a new sub-part of vertex embeddings. The vertex embeddings that have been trained on all sample blocks within this computing node can also be sent to a new node for training on a new set of sample blocks.
\end{itemize}

By tuning the number and size of buffers in our data partitioning strategy conditioned by specific network bandwidth, memory bandwidth, and computing power of CPUs and GPUs, we could find the optimal settings for our system to run most efficiently on a specific GPU cluster.

\section{System Implementation}
\label{sec:impl}
\subsection{Walk Engine Implementation}
We adopt the walk engine in a state-of-the-art distributed graph processing system~\footnote{https://github.com/Tencent/plato} and for our network scale, propose two implementations according to different cluster settings:
\begin{itemize}
  \item Some clusters have a large gap of computing power between GPUs and CPUs as well as relatively slower disk and smaller memory. On these clusters, if we run random walk as an online process before the embedding training, the pipeline will stall. Also, it is impractical to store all the edge samples needed in one epoch in memory. In this case, we divide the random walking process into multiple stages and make it an asynchronous offline process. In the first stage we generate random walks for the whole network and write them into files partitioned by episode. While running one episode, each machine will read random walk data from one specific partition, do the augmentation and send generated samples to its local GPUs for picking up context and vertex embeddings. Our design allows us to choose from various random walk implementations and make arbitrary modifications. We chose a state-of-the-art distributed walk engine~\cite{Yang:2019:KAF} and improved on it with the degree-guided strategy~\cite{Zhu:2019:GVA} while partitioning the generated random walks.
  \item For high-performance GPU clusters with high-end CPUs, large memory, shared storage, and fast disk I/O (e.g. SSD storage), we merge the network augmentation stage into random walking stage, and store the generated edge samples directly into SSDs with memory mapping, so that embedding training engine can directly access them during training.
\end{itemize}

Our decoupled design of the walk engine and embedding training engine provides flexibility of using different computing resources and running settings for our walk engine and our embedding training engine. Specifically, we run our walk engine for the next epoch while embedding training engine trains samples for this epoch. Running settings are fine-tuned so that for one episode, our walk engine uses shorter run time than the embedding training engine, thus can be pipelined to give better end-to-end performance.

\subsection{Two-level Ring-based Communication}
As the bandwidths for different types of communication (inter-node communication, cross-socket inter-GPU communication, and GPU peer-to-peer communication) vary greatly, a simple ring-based communication method that includes all GPUs on all nodes is not performant. Thus we use a two-level strategy: on the node level, each node forms an internal ring that includes all local GPUs on that node; on the cluster level, all nodes form another ring. Since we use pipeline design for embedding training, the slower inter-node communication could be completely hidden (See two light blue block sections in Figure~\ref{fig:pipeline}). Note that since we will transfer vertex embeddings back to CPU for the node-level ring-based communication, we do not use GPU direct RDMA and can still achieve high utilization for both computation and communication.

\subsection{Topology-Aware GPU Communication}
On a typical two-socket node with eight GPUs, all GPUs are divided into two groups: the first four and the last four (as shown in Figure~\ref{fig:hybrid}). To avoid cross-socket inter-GPU communication, which is roughly 30\% slower than inter-GPU communication on the same socket, we improved our ring-based parameter communication strategy.

We first enable peer-access for GPUs within one socket. When we transfer vertex embeddings between GPUs during the embedding training, if we detect that two GPUs are on the same socket, we use peer-to-peer memory copy, if two GPUs are cross-socket (this situation will happen twice for a two-socket node), we pipeline the vertex embedding transfer with one device-to-host memory copy and one host-to-device memory copy. Since most of the intra-node GPU communication is done by peer-to-peer memory copy, we can cut the communication traffic to half when compared to GraphVite, which uses CPU as a parameter server.

\section{Experiments \& Results}
\label{sec:exp}
\subsection{System and Hardware Information}

We have two sets of hardware settings run on CentOS 7.2:
\begin{itemize}
  \item {\textbf{Set A:}} runs on a GPU cluster where each node has two 2.50GHz Intel 24-core Xeon Platinum 8255C processors (with hyper-threading), 364GB of main memory, eight 32GB memory NVIDIA V100 GPUs with NVLink 2.0, and one NVMe SSD storage. All nodes in cluster are connected with 100Gb/s InfiniBand switch.
  \item {\textbf{Set B:}} runs on a GPU cluster where each node has two 2.20GHz Intel 22-core Xeon E5-2699 v4 processors (with hyper-threading), 239GB of main memory, eight NVIDIA P40 GPUs with 24GB on-board memory. All nodes in cluster are connected with 40Gb/s network connections.
\end{itemize}
The distributed program was compiled with GCC 7.3.1, nvcc 10.0, and MPICH-3.2.1.

\subsection{Datasets and Applications}

\begin{table}[!htbp]
\caption{Datasets details for our experiments.}
\begin{center}
\begin{tabular}{|c|c|c|c|}
\hline
\textbf{Dataset}&\multicolumn{2}{|c|}{\textbf{Details}} \\
\cline{2-3} 
\textbf{Name} & \textbf{\textit{nodes}}& \textbf{\textit{edges}}\\
\hline
youtube  & 1,138,499 & 4,945,382 \\
hyperlink-pld & 39,497,204 & 623,056,313 \\
friendster  & 65,608,366 & 1,806,067,135 \\
kron & 2,097,152 & 91,042,010 \\
delaunay & 16,777,216 & 50,331,601 \\
anonymized A & 1,050,000,000 & 280,000,000,000 \\
anonymized B & 1,050,000,000 & 300,000,000,000 \\
generated A & 250,000,000 & 20,000,000,000 \\
generated B & 100,000,000 & 10,000,000,000 \\
\hline

\end{tabular}
\label{tab:dataset}
\end{center}
\end{table}

\begin{table*}[tb]
\caption{Overall performance of our system with one epoch of training on various datasets.}
\begin{center}
\begin{tabular}{|c|c|c|c|c|c|c|c|}
\hline
\textbf{Experiment}&\multicolumn{6}{|c|}{\textbf{Details}}&{\textbf{Time (sec)}} \\
\cline{2-7} 
\textbf{Settings} & \textbf{\textit{framework}} & \textbf{\textit{dateset}} & \textbf{\textit{nodes}} & \textbf{\textit{edges}} & \textbf{\textit{embedding dimension}} & \textbf{\textit{negative samples}} &\\
\hline
8 V100 GPUs & GraphVite & Friendster & 65.6 Million & 1.8 Billion & 96 & 5 & 45.04 \\
8 V100 GPUs & Ours & Friendster & 65.6 Million & 1.8 Billion & 96 & 5 & 3.12 \\
16 V100 GPUs & Ours & Generated B & 100 Million & 10 Billion & 96 & 5 & 15.1 \\
16 V100 GPUs & Ours & Generated A & 250 Million & 20 Billion & 96 & 5 & 27.9 \\
40 V100 GPUs & Ours & Anonymous A & 1.05 Billion & 280 Billion & 128 & 5 & 200 \\
40 P40 GPUs  & Ours & Anonymous B & 1.05 Billion & 300 Billion & 100 & 5 & 1260 \\
\hline
\end{tabular}
\label{tab:perf}
\end{center}
\end{table*}

We use the following datasets for our experiments (details in Table~\ref{tab:dataset}).
\begin{itemize}
  \item {\textbf{YOUTUBE}~\cite{Mislove:2007:MAO}} is a social network dataset crawled from YouTube (www.youtube.com) that consists of over 1.1 million nodes (users) and 4.9 million edges (links);
  \item {\textbf{HYPERLINK-PLD}~\cite{Meusel:2015:TGS}} is a label-free hyperlink network with 43 million nodes and 623 million edges;
  \item {\textbf{FRIENDSTER}~\cite{Yang:2015:DEN}} is a very large social network on a social gaming site with 65 million nodes and 1.8 billion edges. Part of the nodes have labels that represent social groups formed by users;
  \item {\textbf{KRON \& DELAUNAY}} are two networks widely used for benchmarking graph processing systems. Kron is a scale-free network where degree distribution is skew and Delaunay is a mesh network where degree distribution is uniform.
  \item {\textbf{ANONYMOUS}} is a set of networks sampled from multiple internal networks with billion level nodes and hundred-billion level edges. We have anonymized the node and edge information but only preserve the network topology information for the purpose of running the experiments.
  \item {\textbf{GENERATED}} is a set of generated networks with hundred-million level nodes and ten-billion level edges that resemble the topology of real-world social networks. We use this set of networks to test the scalability and performance of our system.
\end{itemize}

Node embeddings can be used in several standard tasks as well as serve as a feature engineering step for downstream machine learning applications. To report the effectiveness of the learned embeddings, we use two tasks that each represents one type of tasks:
\begin{itemize}
  \item {\textbf{Link Prediction}} predicts whether there will be links between two nodes based on the attribute information and the observed existing link information. Following previous works, we use the metric AUC to measure the performance for this task;
  \item {\textbf{Feature Engineering}} uses node embedding on a label-free network to generate node embeddings for an internal downstream machine learning application. We also use the metric AUC to measure the performance for this task.
\end{itemize}

\subsection{Performance Analysis}

\subsubsection{Overall Performance}

Table~\ref{tab:perf} shows the performance of our system on different hardware settings and different networks. Note that since GraphVite is not scalable to multi-node multi-GPU, we only compare with GraphVite on a single-node machine with 8 V100 GPUs. Table~\ref{tab:perf} only includes results on the largest network that can run on GraphVite: Friendster. Table~\ref{tab:scale_compares} shows a more comprehensive single-node performance analysis between our system and GraphVite.

\textbf{Speedup:} On Friendster dataset, we achieve 14.4x speedup over GraphVite when using 8 V100 GPUs. This comes from two parts: 1) our intra-node embedding data partitioning strategy and topology-aware GPU communication implementation reduce both GPU peer-to-peer communication cost as well as host to GPU and GPU to host communication cost, and 2) our pipeline design overlaps most communication process with the computation.

\textbf{Scaling:} Our results on 16 V100 GPUs (two nodes) shows our capability of scaling to cluster that has more nodes when working with larger networks and larger number of embedding dimension. The results on two generated networks show our scalability: on generated B network, which has 0.5x number of edges compared to generated A network, we show a 1.85x speedup.

\textbf{Flexibility:} The results on cluster with V100 GPUs and P40 GPUs show that our system adapts well to different hardware settings and various network sizes with different embedding dimensions. The performance drop of P40 cluster compared to V100 cluster comes from a combination of: 1) device memory bandwidth difference, 2) inter-node network connection difference, 3) random walk generating and storing difference, and 4) FP32 performance difference. Even using P40 cluster, our system only need 35 hours to finish a 100 epoch training of Anonymous B network (its augmented network contains three trillion edges), which is not trainable for any previous embedding training work.

\subsubsection{Evaluation on Node Embedding Tasks}
%accuracy performance (link prediction)
%\begin{figure}[t]
%  \centering
%  \subcaptionbox{YouTube}[.48\linewidth][c]{%
%    \includegraphics[width=\linewidth]{plots/auc_youtube.pdf}}
%  \subcaptionbox{Hyperlink}[.48\linewidth][c]{%
%    \includegraphics[width=\linewidth]{plots/auc_hyper.pdf}}
%  \caption{AUC curve for link prediction evaluation using GraphVite and our system. Under the same number of epoch of training, our system achieves better AUC on YouTube dataset and almost identical AUC on Hyperlink dataset.}
%  \label{fig:auc}
%\end{figure}

\begin{table}[htbp]
\caption{Evaluate AUC of link prediction after 1000 epochs of node embedding training using GraphVite and our system.}
\begin{center}
\begin{tabular}{|c|c|c|}
\hline
\textbf{Dataset}&\textbf{Framework}&\textbf{Final Evaluation AUC}\\
\hline
    {Youtube}&GraphVite & 0.909\\
    &Ours & 0.926 \\
    \hline
    {Hyperlink}&GraphVite & 0.989 \\
    &Ours & 0.988 \\
    \hline
\end{tabular}
\label{tab:final_auc}
\end{center}
\end{table}

\begin{figure*}[htbp]
  \centering
    \centerline{\includegraphics[width=0.7\linewidth]{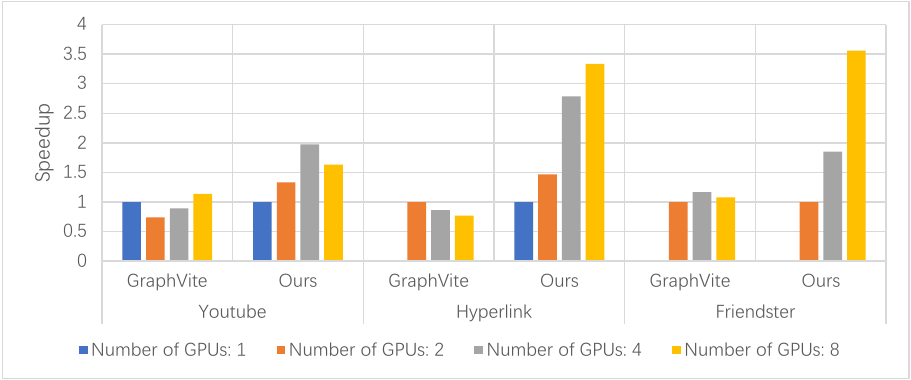}}
  \caption{Intra-node scalability of our system and GraphVite on three datasets.}
  \label{fig:one_node_scale}
\end{figure*}

\begin{itemize}
\item{\textbf{Link Prediction:}} We evaluate link prediction task on two open datasets: YouTube dataset and Hyperlink-PLD dataset. To compare with GraphVite, we adopt its method for link prediction evaluation: We split the edges into three sets: training set, test set and validation set. For training set, the negative samples are generated during the training, for test and validation set, we generate negative samples by randomly picking up node pairs that are not real edges in the network. We use 1\% and 0.01\% edges for test and validation on YouTube and Hyperlink-PLD dataset respectively. We also keep the same training settings as GraphVite, such as the learning rate, the number of negative samples, and embedding initialization method.

Table~\ref{tab:final_auc} presents the performance of GraphVite and our system over different training epochs on these datasets. On YouTube dataset, our system has generally maintained a higher AUC and achieves the best AUC 200 epochs earlier than GraphVite. For Hyperlink-PLD, we have competitive AUC at the end of the training. Since our walk engine is decoupled with our embedding training engine, we could generate random walks of arbitrary epoch sizes, for example in our experiment, we generate random walks for 10 epochs, then repeatedly use these walks to launch a 100-epoch training process. This flexible setting allows us to tune run time of the walk engine so that it can be fully hidden by the embedding training in the pipeline. The results of the experiment show that this extra flexibility does not hurt the final performance of the model.

%compare auc of an internal task
\item{\textbf{Feature Engineering:}} On one anonymous dataset, we compare our embedding system with a CPU node embedding implementation of LINE~\cite{Tang:2015:LINE} after training for the same number of epochs. Without generality, we set this number to 10, which is empirically enough for the model to converge in our internal task. On training AUC, our GPU implementation achieves competitive result compared to the CPU implementation (the difference is within \%0.1). On evaluation AUC, our GPU implementation achieves better result. Further evaluation of the downstream task shows that embeddings generated by our system can achieve the same end-to-end performance as embeddings generated by the CPU implementation.
\end{itemize}

\begin{table}[htbp]
\caption{Evaluate AUC of an internal task using node embedding for feature engineering.}
\begin{center}
\begin{tabular}{|c|c|c|}
\hline
\textbf{Embedding}&\multicolumn{2}{|c|}{\textbf{Details}} \\
\cline{2-3} 
\textbf{Algorithm} & \textbf{\textit{Training AUC}}& \textbf{\textit{Evaluation AUC}} \\
\hline
CPU Embedding  & 0.81147 & 0.79996 \\
GPU Embedding (ours)  & 0.80996 & 0.80008 \\
\hline

\end{tabular}
\label{tab:auc}
\end{center}
\end{table}

\subsubsection{Scalability}

%\begin{figure}[htbp]
%  \centerline{\includegraphics[width=\linewidth]{plots/strong_scale.pdf}}
%  \caption{Scalability of our system running on two large-scale networks.}
%  \label{fig:scalability}
%\end{figure}

%strong/weak scaling: V100 and P40, 2x2 sets.
To test the inter-node scalability of our system, we use two networks that represent typical degree distribution and topology of real-world social networks. The two networks contain 10 billion (generated B) and 25 billion (generated A) edges respectively. We run our embedding training on each network for 10 epochs and record the average run time per epoch, as shown in Table~\ref{tab:scale_ours}. On generated A and generated B, we show 1.67x and 1.85x speedup over one-node 8 GPU setting when using two-node 16 GPU setting. This scalability is achieved through our hierarchical data partition design and various optimizations in our embedding training pipeline implementation, which have maximized the overlap of inter-node vertex embedding communication and intra-node training.

\begin{table}[htbp]
\caption{Scalability on Various Networks.}
\begin{center}
\begin{tabular}{|c|c|c|c|c|c|}
\hline
{\textbf{Dataset}}&\multicolumn{5}{|c|}{\textbf{Number of GPUs (time:sec)}}\\
\cline{2-6}
&\textbf{1}&\textbf{2}&\textbf{4}&\textbf{8}&\textbf{16} \\
\hline
    youtube & 0.16 & 0.12 & 0.081 & 0.098 & N/A\\
    \hline
    hyperlink-pld & 6.6 & 4.5 & 2.37 & 1.98 & N/A\\
    \hline
    friendster & N/A & 11.1 & 6 & 3.12 & N/A\\
    \hline
    kron & 4.6 & 2.8 & 1.46 & 0.75 & N/A\\
    \hline
    delaunay & 2.16 & 1.16 & 0.59 & 0.34 & N/A\\
    \hline
    generated A & N/A & N/A & 30.3 & 27.9 & 15.1 \\
    \hline
    generated B & N/A & N/A & N/A & 59.6 & 35.7 \\
    \hline
\end{tabular}
\label{tab:scale_ours}
\end{center}
\end{table}

We also compare the intra-node scalability of our system and GraphVite. Table~\ref{tab:scale_compares} and Figure~\ref{fig:one_node_scale} show run time of GraphVite and our system with a 10-epoch training session. We record the average per-epoch run time. Our system has both better performance as well as better scalability on 1, 2, 4, and 8 GPUs within a node. On three open datasets, our system achieves 5.9x--14.4x speedup in run time over GraphVite. We also show better scalability as the size of network getting larger. The main reason is that our hybrid model data parallel training has effectively reduced the communication cost than in GraphVite, where CPU is served as the parameter server and no pipeline design is used. Note that to force the same synchronization ratio, we need to adjust episode size in GraphVite with the number of GPUs. We argue that this is the setting that GraphVite should use to enable baseline accuracy and make a fair comparison with our system.

Table~\ref{tab:scale_ours} shows our intra-node scalability on various networks. For small datasets such as YOUTUBE, it is difficult to hide communication cost since the computation workload is too light-weighted, thus leads to a worse scalability.

\begin{table}[htbp]
\caption{Intra-node Scalability Comparison with GraphVite.}
\begin{center}
\begin{tabular}{|c|c|c|c|c|c|}
\hline
{\textbf{Dataset}}&{\textbf{Framework}}&\multicolumn{4}{|c|}{\textbf{Number of GPUs (time:sec)}}\\
\cline{3-6}
&&\textbf{1}&\textbf{2}&\textbf{4}&\textbf{8} \\
\hline
    {Youtube}&GraphVite & 0.66 & 0.89 & 0.74 & 0.58\\
    &Ours & \textbf{0.16} & \textbf{0.12} & \textbf{0.081} & \textbf{0.098} \\
    \hline
    {Hyperlink}&GraphVite & N/A & 19.02 & 22 & 24.71 \\
    &Ours & 6.6 & \textbf{4.5} & \textbf{2.37} & \textbf{1.98} \\
    \hline
    {Friendster}&GraphVite & N/A & 48.51 & 41.51 & 45.04\\
    &Ours & N/A & \textbf{11.1} & \textbf{6} & \textbf{3.12} \\
    \hline
\end{tabular}
\label{tab:scale_compares}
\end{center}
\end{table}

\section{Related Works}
\label{sec:related}
Node embedding can either be used in tasks such as node classification, node clustering, node ranking, and link prediction, or serve as a feature engineering stage in complex network-based machine learning pipelines. Methods for node embedding can roughly be categorized into three types: factorization based methods, deep learning based methods, and random-walk based methods.

Our work follows the direction of the scalable algorithms that built on either edge or path samples of networks and the parallel graph embedding systems.

\subsection{Scalable Node Embedding Algorithms}
Most of node embedding algorithms \cite{Perozzi:2014:DeepWalk,Grover:2016:Node2Vec,Dong:2017:Metapath2Vec} fall into the following paradigm: firstly, generating sequence of nodes, then leveraging language modeling techniques to generate node embeddings.
DeepWalk \cite{Perozzi:2014:DeepWalk} generates node sequences by doing random walks on a given graph, then it applies SkipGram language model \cite{Mikolov:2013:SkipGram} to learn the node embeddings.
Node2Vec \cite{Grover:2016:Node2Vec} proposes a high-order walk strategy that the next walk step depends on the previous one. It can be considered as a combination of Breadth-first Sampling(BFS) and Depth-first Sampling(DFS) strategy.
Metapath2Vec \cite{Dong:2017:Metapath2Vec} is designed for heterogeneous graphs. The generated node sequence follows a \textit{meta-path schema}, which defines the edge type between nodes within a sequence.

\subsection{Parallel Word Embedding Systems}
Most node embedding algorithms leverage word embedding techniques to generate the latent representation of nodes. There are many systems targeting at optimizing word embedding algorithms like word2vec \cite{Mikolov:2013:Word2Vec}.

Ordentlich et al. \cite{Ordentlich:2016:network} proposes a model-parallel approach for distributed word2vec training. Instead of partitioning model at word dimension, it partitions the model at embedding dimension. Each server manages part of embeddings for all words. When performing dot product of two word embeddings, it first computes the partial dot product locally, then performs a reduce operation among servers to get the final results. When the embedding dimension is large, the cost of network transmission can be amortized. The proposed system can scale to datasets that contain several hundred million words on a CPU cluster.
One of the core components in word2vec is negative sampling. Stergiou et al. \cite{Stergiou:2017:distributed} find most of the network transfer comes from the negative samples. They then proposed several negative sampling strategies to reduce the network transmission overhead. Their system can scale to datasets with a vocabulary size of more than one billion on a CPU cluster.
Ji et al. \cite{Ji:2019:parallelizing}  proposed a new parallelization scheme for word2vec computation in shared memory. By forming the computation into mini-batches, they can share the negative samples within one mini-batch. By this way, level-1 BLAS operations can be converted into level-3 BLAS matrix multiply operations. It also achieves good strong-scalability on CPUs.
BlazingText \cite{Gupta:2017:blazingtext} is a multi-GPU word embedding system. It adopts a similar idea of sharing negative samples inside each batch. Instead of using cuBLAS kernels, it proposed customized GPU kernels to reduce kernel launch overhead. For distributed training, it uses data-parallel approach, word embeddings are replicated on each GPU.  BlazingText achieves near-linear scalability on 8 GPUs. But it does not apply to billion-scale word embedding problems, because it lacks the capability of handling embeddings that do not fit into GPU memory.

\subsection{Parallel Graph Embedding Systems}
Several deep learning frameworks support multi-node multi-GPU training~\cite{Abadi:2015:TF,Paszke:2019:PT,Chen:2015:MXNET}. However, none of the existing deep learning frameworks offer: 1) native data structures or primitives for doing network-related operations such as random walk; 2) a hybrid model data parallel training execution model to easily handle applications such as large-scale node embedding.

PyTorch-BigGraph~\cite{Lerer:2019:PBG} (PBG) is a scalable knowledge graph embedding framework built on top of PyTorch. It randomly partitions the graph into 2-D blocks. Blocks can be stored into either CPU memory or external storage so that the system can scale to arbitrarily large graphs. DGL-KE~\cite{Zheng:2020:DGLKE} is another distributed knowledge graph embedding framework. It has similar idea with Pytorch-BigGraph. It supports distributed multi-gpu training. Different from PBG, it uses graph partition algorithm in METIS~\cite{Karypis:1998:METIS} to reduce network overhead. It also proposes a new negative sampling strategy to improve data locality. Both PBG and DGL-KE target at knowledge graph embedding, which does not contain random-walk-based network augmentation and usually does not operate on large-scale networks with billion level nodes.

C. Bayan Brudss et al.~\cite{BayanBruss:2019:GES} has developed a scalable graph embedding system that could run on a heterogeneous cluster. They adopt the parameter server (PS) design~\cite{Li:2014:SDM} and use embedding lookup operations to asynchronously train/update the model. This introduces large communication overhead between parameter servers and workers and makes it difficult to scale to a lot of real-world networks at larger scale. Also, their asynchronous training strategy is sensitive to the number of workers, fails to provide the best accuracy with just enough computing resources to train the model.

GraphVite~\cite{Zhu:2019:GVA} is a general graph embedding engine that runs on a multi-GPU single-node machine. It is the first of its kind to explore scalable node embedding using GPUs and has largely improved on the training time for medium-size node embedding tasks. However, its design uses CPU as a parameter server to run random walk online and transfer embeddings between GPUs, also the system lacks a pipeline design to properly overlap communication with computation, which makes it difficult to scale to large-scale node embedding on multi-node cluster without a major change in system design.

\section{Conclusion and Future Work}
\label{sec:conc}
We present a high-performance system that uses hybrid model data parallel training and runs on a distributed GPU cluster to handle the challenging large-scale network node embedding problem. We develop hierarchical data partitioning strategy and design an embedding training pipeline to maximize the overlap between communication and computation. Our design decouples the random walk engine that runs network augmentation with the embedding training engine. Our system is highly scalable to multi-GPU multi-node GPU clusters and achieves an order of magnitude speedup on average over the current state-of-the-art GPU node embedding system with competitive or better accuracy on open datasets. We expect our system and its design decisions will directly impact the development of future large-scale GPU network embedding frameworks. In the future, we plan to generalize our system to more types of graph representation learning and knowledge graph embedding models on large-scale networks with various tasks and applications.

\section*{Acknowledgment}
We appreciate the technical assistance, advice and machine access from colleagues at Tencent: Chong (Tiller) Zha, Shiyong (Warrior) Wang, Haidong (Hudson) Rong, and Zhiheng (Nickey) Lin. We also thank Stanley Tzeng for his proofreading of the manuscript.

\bibliographystyle{bib/IEEEtran}
\bibliography{bib/IEEEabrv,bib/main}

\begin{thebibliography}{10}
\providecommand{\url}[1]{#1}
\csname url@samestyle\endcsname
\providecommand{\newblock}{\relax}
\providecommand{\bibinfo}[2]{#2}
\providecommand{\BIBentrySTDinterwordspacing}{\spaceskip=0pt\relax}
\providecommand{\BIBentryALTinterwordstretchfactor}{4}
\providecommand{\BIBentryALTinterwordspacing}{\spaceskip=\fontdimen2\font plus
\BIBentryALTinterwordstretchfactor\fontdimen3\font minus
  \fontdimen4\font\relax}
\providecommand{\BIBforeignlanguage}[2]{{%
\expandafter\ifx\csname l@#1\endcsname\relax
\typeout{** WARNING: IEEEtran.bst: No hyphenation pattern has been}%
\typeout{** loaded for the language `#1'. Using the pattern for}%
\typeout{** the default language instead.}%
\else
\language=\csname l@#1\endcsname
\fi
#2}}
\providecommand{\BIBdecl}{\relax}
\BIBdecl

\bibitem{Lerer:2019:PBG}
A.~Lerer, L.~Wu, J.~Shen, T.~Lacroix, L.~Wehrstedt, A.~Bose, and
  A.~Peysakhovich, ``{PyTorch-BigGraph: A Large-scale Graph Embedding
  System},'' in \emph{Proceedings of the 2nd SysML Conference}, Palo Alto, CA,
  USA, 2019.

\bibitem{Wang:2018:BSC}
\BIBentryALTinterwordspacing
J.~Wang, P.~Huang, H.~Zhao, Z.~Zhang, B.~Zhao, and D.~L. Lee, ``Billion-scale
  commodity embedding for e-commerce recommendation in alibaba,'' in
  \emph{Proceedings of the 24th ACM SIGKDD International Conference on
  Knowledge Discovery and Data Mining}, ser. KDD ’18.\hskip 1em plus 0.5em
  minus 0.4em\relax New York, NY, USA: Association for Computing Machinery,
  2018, p. 839–848. [Online]. Available:
  \url{https://doi.org/10.1145/3219819.3219869}
\BIBentrySTDinterwordspacing

\bibitem{Zhu:2019:GVA}
\BIBentryALTinterwordspacing
Z.~Zhu, S.~Xu, J.~Tang, and M.~Qu, ``Graphvite: A high-performance cpu-gpu
  hybrid system for node embedding,'' in \emph{The World Wide Web Conference},
  ser. WWW ’19.\hskip 1em plus 0.5em minus 0.4em\relax New York, NY, USA:
  Association for Computing Machinery, 2019, p. 2494–2504. [Online].
  Available: \url{https://doi.org/10.1145/3308558.3313508}
\BIBentrySTDinterwordspacing

\bibitem{Tang:2015:LINE}
\BIBentryALTinterwordspacing
J.~Tang, M.~Qu, M.~Wang, M.~Zhang, J.~Yan, and Q.~Mei, ``Line: Large-scale
  information network embedding,'' in \emph{Proceedings of the 24th
  International Conference on World Wide Web}, ser. WWW ’15.\hskip 1em plus
  0.5em minus 0.4em\relax Republic and Canton of Geneva, CHE: International
  World Wide Web Conferences Steering Committee, 2015, p. 1067–1077.
  [Online]. Available: \url{https://doi.org/10.1145/2736277.2741093}
\BIBentrySTDinterwordspacing

\bibitem{Yang:2021:EAE}
D.~{Yang}, J.~{Liu}, and J.~{Lai}, ``Edges: An efficient distributed graph
  embedding system on gpu clusters,'' \emph{IEEE Transactions on Parallel and
  Distributed Systems}, vol.~32, no.~7, pp. 1892--1902, 2021.

\bibitem{dutta:2018:STALEGRAD}
S.~Dutta, G.~Joshi, S.~Ghosh, P.~Dube, and P.~Nagpurkar, ``Slow and stale
  gradients can win the race: Error-runtime trade-offs in distributed sgd,''
  2018.

\bibitem{Yang:2019:KAF}
\BIBentryALTinterwordspacing
K.~Yang, M.~Zhang, K.~Chen, X.~Ma, Y.~Bai, and Y.~Jiang, ``Knightking: A fast
  distributed graph random walk engine,'' in \emph{Proceedings of the 27th ACM
  Symposium on Operating Systems Principles}, ser. SOSP ’19.\hskip 1em plus
  0.5em minus 0.4em\relax New York, NY, USA: Association for Computing
  Machinery, 2019, p. 524–537. [Online]. Available:
  \url{https://doi.org/10.1145/3341301.3359634}
\BIBentrySTDinterwordspacing

\bibitem{Mislove:2007:MAO}
\BIBentryALTinterwordspacing
A.~Mislove, M.~Marcon, K.~P. Gummadi, P.~Druschel, and B.~Bhattacharjee,
  ``Measurement and analysis of online social networks,'' in \emph{Proceedings
  of the 7th ACM SIGCOMM Conference on Internet Measurement}, ser. IMC
  ’07.\hskip 1em plus 0.5em minus 0.4em\relax New York, NY, USA: Association
  for Computing Machinery, 2007, p. 29–42. [Online]. Available:
  \url{https://doi.org/10.1145/1298306.1298311}
\BIBentrySTDinterwordspacing

\bibitem{Meusel:2015:TGS}
\BIBentryALTinterwordspacing
R.~Meusel, S.~Vigna, O.~Lehmberg, and C.~Bizer, ``The graph structure in the
  web – analyzed on different aggregation levels,'' \emph{The Journal of Web
  Science}, vol.~1, no.~1, pp. 33--47, 2015. [Online]. Available:
  \url{http://dx.doi.org/10.1561/106.00000003}
\BIBentrySTDinterwordspacing

\bibitem{Yang:2015:DEN}
\BIBentryALTinterwordspacing
J.~Yang and J.~Leskovec, ``Defining and evaluating network communities based on
  ground-truth,'' \emph{Knowl. Inf. Syst.}, vol.~42, no.~1, p. 181–213, Jan.
  2015. [Online]. Available: \url{https://doi.org/10.1007/s10115-013-0693-z}
\BIBentrySTDinterwordspacing

\bibitem{Perozzi:2014:DeepWalk}
B.~Perozzi, R.~Al-Rfou, and S.~Skiena, ``Deepwalk: Online learning of social
  representations,'' in \emph{Proceedings of the 20th ACM SIGKDD international
  conference on Knowledge discovery and data mining}, 2014, pp. 701--710.

\bibitem{Grover:2016:Node2Vec}
A.~Grover and J.~Leskovec, ``node2vec: Scalable feature learning for
  networks,'' in \emph{Proceedings of the 22nd ACM SIGKDD international
  conference on Knowledge discovery and data mining}, 2016, pp. 855--864.

\bibitem{Dong:2017:Metapath2Vec}
Y.~Dong, N.~V. Chawla, and A.~Swami, ``metapath2vec: Scalable representation
  learning for heterogeneous networks,'' in \emph{Proceedings of the 23rd ACM
  SIGKDD international conference on knowledge discovery and data mining},
  2017, pp. 135--144.

\bibitem{Mikolov:2013:SkipGram}
T.~Mikolov, K.~Chen, G.~Corrado, and J.~Dean, ``Efficient estimation of word
  representations in vector space,'' \emph{arXiv preprint arXiv:1301.3781},
  2013.

\bibitem{Mikolov:2013:Word2Vec}
T.~Mikolov, I.~Sutskever, K.~Chen, G.~S. Corrado, and J.~Dean, ``Distributed
  representations of words and phrases and their compositionality,'' in
  \emph{Advances in neural information processing systems}, 2013, pp.
  3111--3119.

\bibitem{Ordentlich:2016:network}
E.~Ordentlich, L.~Yang, A.~Feng, P.~Cnudde, M.~Grbovic, N.~Djuric,
  V.~Radosavljevic, and G.~Owens, ``Network-efficient distributed word2vec
  training system for large vocabularies,'' in \emph{Proceedings of the 25th
  ACM International on Conference on Information and Knowledge Management},
  2016, pp. 1139--1148.

\bibitem{Stergiou:2017:distributed}
S.~Stergiou, Z.~Straznickas, R.~Wu, and K.~Tsioutsiouliklis, ``Distributed
  negative sampling for word embeddings,'' in \emph{Thirty-First AAAI
  Conference on Artificial Intelligence}, 2017.

\bibitem{Ji:2019:parallelizing}
S.~Ji, N.~Satish, S.~Li, and P.~K. Dubey, ``Parallelizing word2vec in shared
  and distributed memory,'' \emph{IEEE Transactions on Parallel and Distributed
  Systems}, vol.~30, no.~9, pp. 2090--2100, 2019.

\bibitem{Gupta:2017:blazingtext}
S.~Gupta and V.~Khare, ``Blazingtext: Scaling and accelerating word2vec using
  multiple gpus,'' in \emph{Proceedings of the Machine Learning on HPC
  Environments}, 2017, pp. 1--5.

\bibitem{Abadi:2015:TF}
\BIBentryALTinterwordspacing
M.~Abadi, A.~Agarwal, P.~Barham, E.~Brevdo, Z.~Chen, C.~Citro, G.~S. Corrado,
  A.~Davis, J.~Dean, M.~Devin, S.~Ghemawat, I.~Goodfellow, A.~Harp, G.~Irving,
  M.~Isard, Y.~Jia, R.~Jozefowicz, L.~Kaiser, M.~Kudlur, J.~Levenberg,
  D.~Man\'{e}, R.~Monga, S.~Moore, D.~Murray, C.~Olah, M.~Schuster, J.~Shlens,
  B.~Steiner, I.~Sutskever, K.~Talwar, P.~Tucker, V.~Vanhoucke, V.~Vasudevan,
  F.~Vi\'{e}gas, O.~Vinyals, P.~Warden, M.~Wattenberg, M.~Wicke, Y.~Yu, and
  X.~Zheng, ``{TensorFlow}: Large-scale machine learning on heterogeneous
  systems,'' 2015, software available from tensorflow.org. [Online]. Available:
  \url{http://tensorflow.org/}
\BIBentrySTDinterwordspacing

\bibitem{Paszke:2019:PT}
\BIBentryALTinterwordspacing
A.~Paszke, S.~Gross, F.~Massa, A.~Lerer, J.~Bradbury, G.~Chanan, T.~Killeen,
  Z.~Lin, N.~Gimelshein, L.~Antiga, A.~Desmaison, A.~Kopf, E.~Yang, Z.~DeVito,
  M.~Raison, A.~Tejani, S.~Chilamkurthy, B.~Steiner, L.~Fang, J.~Bai, and
  S.~Chintala, ``Pytorch: An imperative style, high-performance deep learning
  library,'' in \emph{Advances in Neural Information Processing Systems 32},
  H.~Wallach, H.~Larochelle, A.~Beygelzimer, F.~d\textquotesingle
  Alch\'{e}-Buc, E.~Fox, and R.~Garnett, Eds.\hskip 1em plus 0.5em minus
  0.4em\relax Curran Associates, Inc., 2019, pp. 8024--8035. [Online].
  Available:
  \url{http://papers.neurips.cc/paper/9015-pytorch-an-imperative-style-high-performance-deep-learning-library.pdf}
\BIBentrySTDinterwordspacing

\bibitem{Chen:2015:MXNET}
\BIBentryALTinterwordspacing
T.~Chen, M.~Li, Y.~Li, M.~Lin, N.~Wang, M.~Wang, T.~Xiao, B.~Xu, C.~Zhang, and
  Z.~Zhang, ``Mxnet: A flexible and efficient machine learning library for
  heterogeneous distributed systems,'' 2015, cite arxiv:1512.01274Comment: In
  Neural Information Processing Systems, Workshop on Machine Learning Systems,
  2016. [Online]. Available: \url{http://arxiv.org/abs/1512.01274}
\BIBentrySTDinterwordspacing

\bibitem{Zheng:2020:DGLKE}
D.~Zheng, X.~Song, C.~Ma, Z.~Tan, Z.~Ye, J.~Dong, H.~Xiong, Z.~Zhang, and
  G.~Karypis, ``Dgl-ke: Training knowledge graph embeddings at scale,'' 2020.

\bibitem{Karypis:1998:METIS}
G.~Karypis and V.~Kumar, ``A fast and high quality multilevel scheme for
  partitioning irregular graphs,'' \emph{SIAM J. Sci. Comput.}, vol.~20, no.~1,
  p. 359–392, Dec. 1998.

\bibitem{BayanBruss:2019:GES}
C.~B. Bruss, A.~Khazane, J.~Rider, R.~Serpe, S.~Nagrecha, and K.~Hines, ``Graph
  embeddings at scale,'' in \emph{Proceedings of the 15th International
  Workshop on Mining and Learning with Graphs (MGL)}, 2019.

\bibitem{Li:2014:SDM}
M.~Li, D.~G. Andersen, J.~W. Park, A.~J. Smola, A.~Ahmed, V.~Josifovski,
  J.~Long, E.~J. Shekita, and B.-Y. Su, ``Scaling distributed machine learning
  with the parameter server,'' in \emph{11th {USENIX} Symposium on Operating
  Systems Design and Implementation ({OSDI} 14)}.\hskip 1em plus 0.5em minus
  0.4em\relax Broomfield, CO: {USENIX} Association, Oct. 2014, pp. 583--598.

\end{thebibliography}

\end{document}